\documentstyle[aps,twocolumn]{revtex}
\begin{document}

%
%
%
\edef\catcodeat{\the\catcode`\@ }     \catcode`\@=11
\newbox\p@ctbox                       
\newbox\t@mpbox                       
\newbox\@uxbox                        
\newbox\s@vebox                       
\newtoks\desct@ks \desct@ks={}        
\newtoks\@ppenddesc                   
\newtoks\sh@petoks                    
\newif\ifallfr@med  \allfr@medfalse   
\newif\if@ddedrows                    
\newif\iffirstp@ss  \firstp@ssfalse   
\newif\if@mbeeded                     
\newif\ifpr@cisebox                   
\newif\ifvt@p                         
\newif\ifvb@t                         
\newif\iff@nished    \f@nishedtrue    
\newif\iffr@med                       
\newif\ifj@stbox     \j@stboxfalse    
\newcount\helpc@unt                   
\newcount\p@ctpos                     
\newdimen\r@leth      \r@leth=0.4pt   
\newdimen\x@nit                       
\newdimen\y@nit                       
\newdimen\xsh@ft                      
\newdimen\ysh@ft                      
\newdimen\@uxdimen                    
\newdimen\t@mpdimen                   
\newdimen\t@mpdimeni                  
\newdimen\b@tweentandp                
\newdimen\b@ttomedge                  
\newdimen\@pperedge                   
\newdimen\@therside                   
\newdimen\d@scmargin                  
\newdimen\p@ctht                      
\newdimen\l@stdepth
\newdimen\in@tdimen
\newdimen\l@nelength
\newdimen\re@lpictwidth
\def\justframes{\global\j@stboxtrue}  
\def\picturemargins#1#2{\b@tweentandp=#1\@therside=#2\relax}
\def\allframed{\global\allfr@medtrue} 
\def\emptyplace#1#2{\pl@cedefs        
    \setbox\@uxbox=\vbox to#2{\n@llpar
        \hsize=#1\vfil \vrule height0pt width\hsize}
    \e@tmarks}
\def\boxplace{\pl@cedefs\afterassignment\re@dvbox\let\n@xt= }
\def\re@dvbox{\setbox\@uxbox=\vbox\bgroup
         \n@llpar\aftergroup\e@tmarks}
\def\fontcharplace#1#2{\pl@cedefs     
    \setbox\@uxbox=\hbox{#1\char#2\/}%
    \xsh@ft=-\wd\@uxbox               
    \setbox\@uxbox=\hbox{#1\char#2}%
    \advance\xsh@ft by \wd\@uxbox     
    \helpc@unt=#2
    \advance\helpc@unt by -63         
    \x@nit=\fontdimen\helpc@unt#1%
    \advance\helpc@unt by  20         
    \y@nit=\fontdimen\helpc@unt#1%
    \advance\helpc@unt by  20         
    \ifnum\helpc@unt<51
      \ysh@ft=-\fontdimen\helpc@unt#1%
    \fi
    \e@tmarks}
\def\n@llpar{\parskip0pt \parindent0pt
    \leftskip=0pt \rightskip=0pt
    \everypar={}}
\def\pl@cedefs{\xsh@ft=0pt\ysh@ft=0pt}
\def\e@tmarks#1{\setbox\@uxbox=\vbox{ 
      \n@llpar
      \hsize=\wd\@uxbox               
      \noindent\copy\@uxbox           
      \kern-\wd\@uxbox                
      #1\par}
    \st@redescription}
\def\t@stprevpict#1{\ifvoid#1\else    
   \errmessage{Previous picture is not finished yet.}\fi} 

\def\st@redescription#1\par{
    \global\setbox\s@vebox=\vbox{\box\@uxbox\unvbox\s@vebox}%
    \desct@ks=\expandafter{\the\desct@ks#1\@ndtoks}}
\def\def@ultdefs{\p@ctpos=1         
      \def\lines@bove{0}
      \@ddedrowsfalse               
      \@mbeededfalse                
      \pr@ciseboxfalse
      \vt@pfalse                    
      \vb@tfalse                    
      \@ppenddesc={}
      \ifallfr@med\fr@medtrue\else\fr@medfalse\fi
      }

\def\descriptionmargins#1{\global\d@scmargin=#1\relax}
\def\@dddimen#1#2{\t@mpdimen=#1\advance\t@mpdimen by#2#1=\t@mpdimen}
\def\placemark#1#2 #3 #4 #5 {\unskip    
      \setbox1=\hbox{\kern\d@scmargin#5\kern\d@scmargin}
      \@dddimen{\ht1}\d@scmargin        
      \@dddimen{\dp1}\d@scmargin        
      \ifx#1l\dimen3=0pt\else           
        \ifx#1c\dimen3=-0.5\wd1\else
          \ifx#1r\dimen3=-\wd1
     \fi\fi\fi
     \ifx#2u\dimen4=-\ht1\else          
       \ifx#2c\dimen4=-0.5\ht1\advance\dimen4 by 0.5\dp1\else
         \ifx#2b\dimen4=0pt\else
           \ifx#2l\dimen4=\dp1
     \fi\fi\fi\fi
     \advance\dimen3 by #3
     \advance\dimen4 by #4
     \advance\dimen4 by-\dp1
     \advance\dimen3 by \xsh@ft         
     \advance\dimen4 by \ysh@ft         
     \kern\dimen3\vbox to 0pt{\vss\copy1\kern\dimen4}
     \kern-\wd1                        
     \kern-\dimen3                     
     \ignorespaces}                    
\def\fontmark #1#2 #3 #4 #5 {\placemark #1#2 #3\x@nit{} #4\y@nit{} {#5} }
\def\fr@msavetopict{\global\setbox\s@vebox=\vbox{\unvbox\s@vebox
      \global\setbox\p@ctbox=\lastbox}%
    \expandafter\firstt@ks\the\desct@ks\st@ptoks}
\def\firstt@ks#1\@ndtoks#2\st@ptoks{%
    \global\desct@ks={#2}%
    \def\t@mpdef{#1}%
    \@ppenddesc=\expandafter\expandafter\expandafter
                        {\expandafter\t@mpdef\the\@ppenddesc}}
\def\testf@nished{{\let\s@tparshape=\relax
    \s@thangindent}}
\def\inspicture{\t@stprevpict\p@ctbox
    \def@ultdefs                  
    \fr@msavetopict
    \iff@nished\else\testf@nished\fi
    \iff@nished\else
      \immediate\write16{Previes picture is not finished yet}%
    \fi
    \futurelet\N@xt\t@stoptions}  
\def\t@stoptions{\let\n@xt\@neletter
  \ifx\N@xt l\p@ctpos=0\else                
   \ifx\N@xt c\p@ctpos=1\else               
    \ifx\N@xt r\p@ctpos=2\else              
     \ifx\N@xt(\let\n@xt\e@tline\else        
      \ifx\N@xt!\@mbeededtrue\else           
       \ifx\N@xt|\fr@medtrue\else            
        \ifx\N@xt^\vt@ptrue\vb@tfalse\else  
         \ifx\N@xt_\vb@ttrue\vt@pfalse\else 
          \ifx\N@xt\bgroup\let\n@xt\@ddgrouptodesc\else
           \let\n@xt\@dddescription 
  \fi\fi\fi\fi\fi\fi\fi\fi\fi\n@xt}
\def\e@tline(#1){\def\lines@bove{#1}
    \@ddedrowstrue
    \futurelet\N@xt\t@stoptions}
\def\@neletter#1{\futurelet\N@xt\t@stoptions} 
\def\@ddgrouptodesc#1{\@ppenddesc={#1}\futurelet\N@xt\t@stoptions}
\def\fr@medpict{\setbox\p@ctbox=
    \vbox{\n@llpar\hsize=\wd\p@ctbox
       \iffr@med\else\r@leth=0pt\fi
       \ifj@stbox\r@leth=0.4pt\fi
       \hrule height\r@leth \kern-\r@leth
       \vrule height\ht\p@ctbox depth\dp\p@ctbox width\r@leth \kern-\r@leth
       \ifj@stbox\hfill\else\copy\p@ctbox\fi
       \kern-\r@leth\vrule width\r@leth\par
       \kern-\r@leth \hrule height\r@leth}}
\def\@dddescription{\fr@medpict     
    \re@lpictwidth=\the\wd\p@ctbox
    \advance\re@lpictwidth by\@therside
    \advance\re@lpictwidth by\b@tweentandp
    \ifhmode\ifinner\pr@ciseboxtrue\fi\fi
    \createp@ctbox
    \let\N@xt\tr@toplacepicture
    \ifhmode                         
      \ifinner\let\N@xt\justc@py
      \else\let\N@xt\vjustc@py
      \fi
    \else
      \ifnum\p@ctpos=1               
        \let\N@xt\justc@py
      \fi
    \fi
    \if@mbeeded\let\N@xt\justc@py\fi 
    \firstp@sstrue
    \N@xt}
\def\createp@ctbox{\global\p@ctht=\ht\p@ctbox
    \advance\p@ctht by\dp\p@ctbox
    \advance\p@ctht by 6pt
    \setbox\p@ctbox=\vbox{
      \n@llpar                     
      \t@mpdimen=\@therside          
      \t@mpdimeni=\hsize             
      \advance\t@mpdimeni by -\@therside
      \advance\t@mpdimeni by -\wd\p@ctbox
      \ifpr@cisebox
        \hsize=\wd\p@ctbox
      \else
        \ifcase\p@ctpos
               \leftskip=\t@mpdimen    \rightskip=\t@mpdimeni
        \or    \advance\t@mpdimeni by \@therside
               \leftskip=0.5\t@mpdimeni \rightskip=\leftskip
        \or    \leftskip=\t@mpdimeni   \rightskip=\t@mpdimen
        \fi
      \fi
      \hrule height0pt             
      \kern6pt                     
      \penalty10000
      \noindent\copy\p@ctbox\par     
      \kern3pt                       
      \hrule height0pt
      \hbox{}%
      \penalty10000
      \interlinepenalty=10000
      \the\@ppenddesc\par            
      \penalty10000                  
      \kern3pt                       
      }%
      \ifvt@p
       \setbox\p@ctbox=\vtop{\unvbox\p@ctbox}%
      \else
        \ifvb@t\else
          \@uxdimen=\ht\p@ctbox
          \advance\@uxdimen by -\p@ctht
          {\vfuzz=\maxdimen
           \global\setbox\p@ctbox=\vbox to\p@ctht{\unvbox\p@ctbox}%
          }%
          \dp\p@ctbox=\@uxdimen
        \fi
      \fi
      }
\def\picname#1{\unskip\setbox\@uxbox=\hbox{\bf\ignorespaces#1\unskip\ }%
      \hangindent\wd\@uxbox\hangafter1\noindent\box\@uxbox\ignorespaces}
\def\justc@py{\ifinner\box\p@ctbox\else\kern\parskip\unvbox\p@ctbox\fi
  \global\setbox\p@ctbox=\box\voidb@x}
\def\vjustc@py{\vadjust{\kern0.5\baselineskip\unvbox\p@ctbox}%
      \global\setbox\p@ctbox=\box\voidb@x}
\def\tr@toplacepicture{
      \ifvmode\l@stdepth=\prevdepth  
      \else   \l@stdepth=0pt         
      \fi
      \vrule height.85em width0pt\par
      \r@memberdims                  
      \global\t@mpdimen=\pagetotal
      \t@stheightofpage              
      \ifdim\b@ttomedge<\pagegoal    
         \let\N@xt\f@gurehere        
         \global\everypar{}
      \else
         \let\N@xt\relax             
         \penalty10000
         \vskip-\baselineskip        
         \vskip-\parskip             
         \immediate\write16{Picture will be shifted down.}%
         \global\everypar{\sw@tchingpass}
      \fi
      \penalty10000
      \N@xt}
\def\sw@tchingpass{
    \iffirstp@ss                     
      \let\n@xt\relax
      \firstp@ssfalse                
    \else
      \let\n@xt\tr@toplacepicture
      \firstp@sstrue
    \fi  \n@xt}
\def\r@memberdims{\global\in@tdimen=0pt
    \ifnum\p@ctpos=0
        \global\in@tdimen=\re@lpictwidth
      \fi
      \global\l@nelength=\hsize
      \global\advance\l@nelength by-\re@lpictwidth
      }
\def\t@stheightofpage{%
     \global\@pperedge=\t@mpdimen
     \advance\t@mpdimen by-0.7\baselineskip 
     \advance\t@mpdimen by \lines@bove\baselineskip 
     \advance\t@mpdimen by \ht\p@ctbox      
     \advance\t@mpdimen by \dp\p@ctbox      
     \advance\t@mpdimen by-0.3\baselineskip 
     \global\b@ttomedge=\t@mpdimen          
     }
\def\f@gurehere{\global\f@nishedfalse
      \t@mpdimen=\lines@bove\baselineskip   
      \advance\t@mpdimen-0.7\baselineskip   
      \kern\t@mpdimen
      \advance\t@mpdimen by\ht\p@ctbox
      \advance\t@mpdimen by\dp\p@ctbox
      {\t@mpdimeni=\baselineskip
       \offinterlineskip
       \unvbox\p@ctbox
       \global\setbox\p@ctbox=\box\voidb@x
       \penalty10000   \kern-\t@mpdimen     
       \penalty10000   \vskip-\parskip      
       \kern-\t@mpdimeni                    
      }%
      \penalty10000                         
      \global\everypar{\s@thangindent}
      }
\def\s@thangindent{%
    \ifdim\pagetotal>\b@ttomedge\global\everypar{}%
      \global\f@nishedtrue             
      \else
        \advance\@pperedge by -1.2\baselineskip
        \ifdim\@pperedge>\pagetotal\global\everypar{}%
          \global\f@nishedtrue
        \else
          \s@tparshape                 
        \fi
        \advance\@pperedge by 1.2\baselineskip
      \fi}
\def\s@tparshape{\t@mpdimen=-\pagetotal
   \advance\t@mpdimen by\b@ttomedge    
   \divide\t@mpdimen by\baselineskip   
   \helpc@unt=\t@mpdimen               
   \advance \helpc@unt by 2            
   \sh@petoks=\expandafter{\the\helpc@unt\space}
   \t@mpdimeni=\lines@bove\baselineskip
   \t@mpdimen=\pagetotal
   \gdef\lines@bove{0}
   \loop \ifdim\t@mpdimeni>0.999\baselineskip 
     \advance\t@mpdimen  by \baselineskip
     \advance\t@mpdimeni by-\baselineskip
     \sh@petoks=\expandafter{\the\sh@petoks 0pt \the\hsize}%
   \repeat
   \loop \ifdim\b@ttomedge>\t@mpdimen         
     \advance\t@mpdimen by \baselineskip
     \sh@petoks=\expandafter{\the\sh@petoks \in@tdimen \l@nelength }%
   \repeat
   \sh@petoks=\expandafter
      {\the\sh@petoks 0pt \the\hsize}
   \expandafter\parshape\the\sh@petoks
   }

\descriptionmargins{2pt}
\picturemargins{15pt}{0pt}

\catcode`\@=\catcodeat        \let\catcodeat=\undefined

\emptyplace{3.3in
\includegraphics{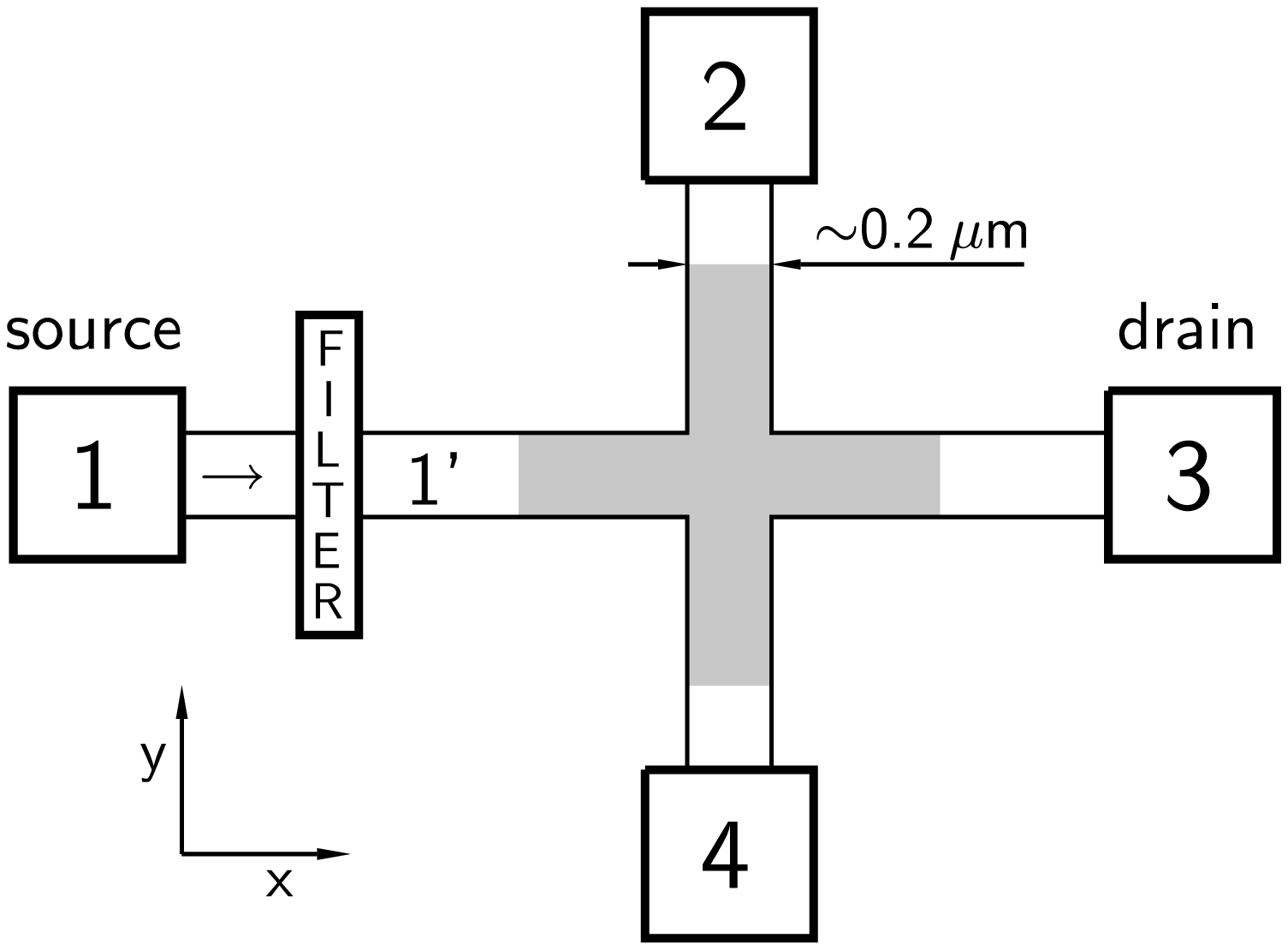}}
{2.6in}
{\footnotesize \noindent
FIG.1. The cross-junction device. Spin-orbit coupling is supposed
to be non-zero in shadowed area only.}

\emptyplace{3.3in
\includegraphics{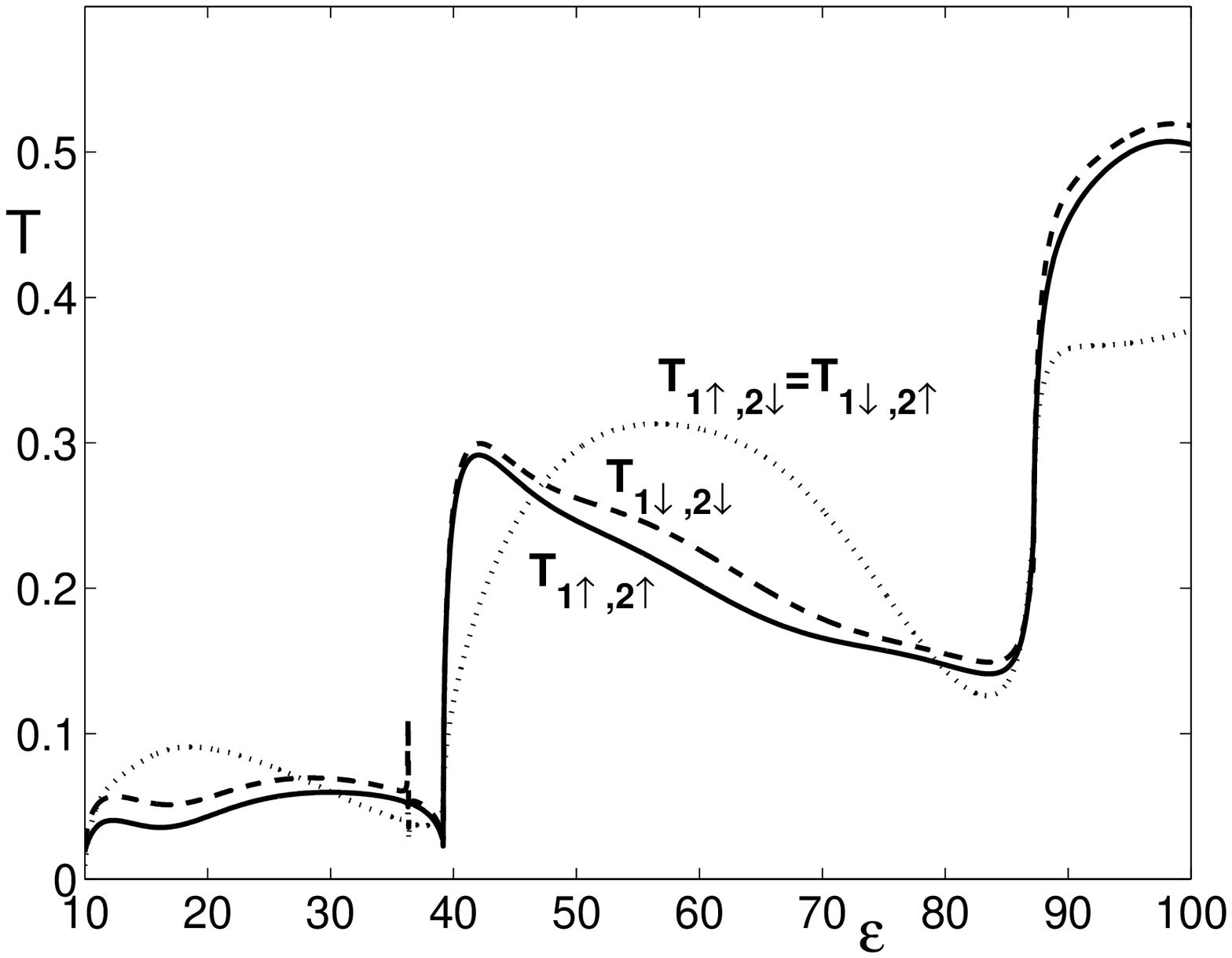}}
{3.1in}
{\footnotesize \noindent
FIG.2. Energy dependence of the partial transmission coefficients
$T_{1 \uparrow ,2 \uparrow}$,  $T_{1 \uparrow ,2 \downarrow}$,
$T_{1 \downarrow ,2 \uparrow}$, and $T_{1 \downarrow ,2 \downarrow}$
describing transition of polarized electron wave incoming  from the
lead 1 into the spin-up and spin-down channels of the lead 2.}

\emptyplace{3.3in
\includegraphics{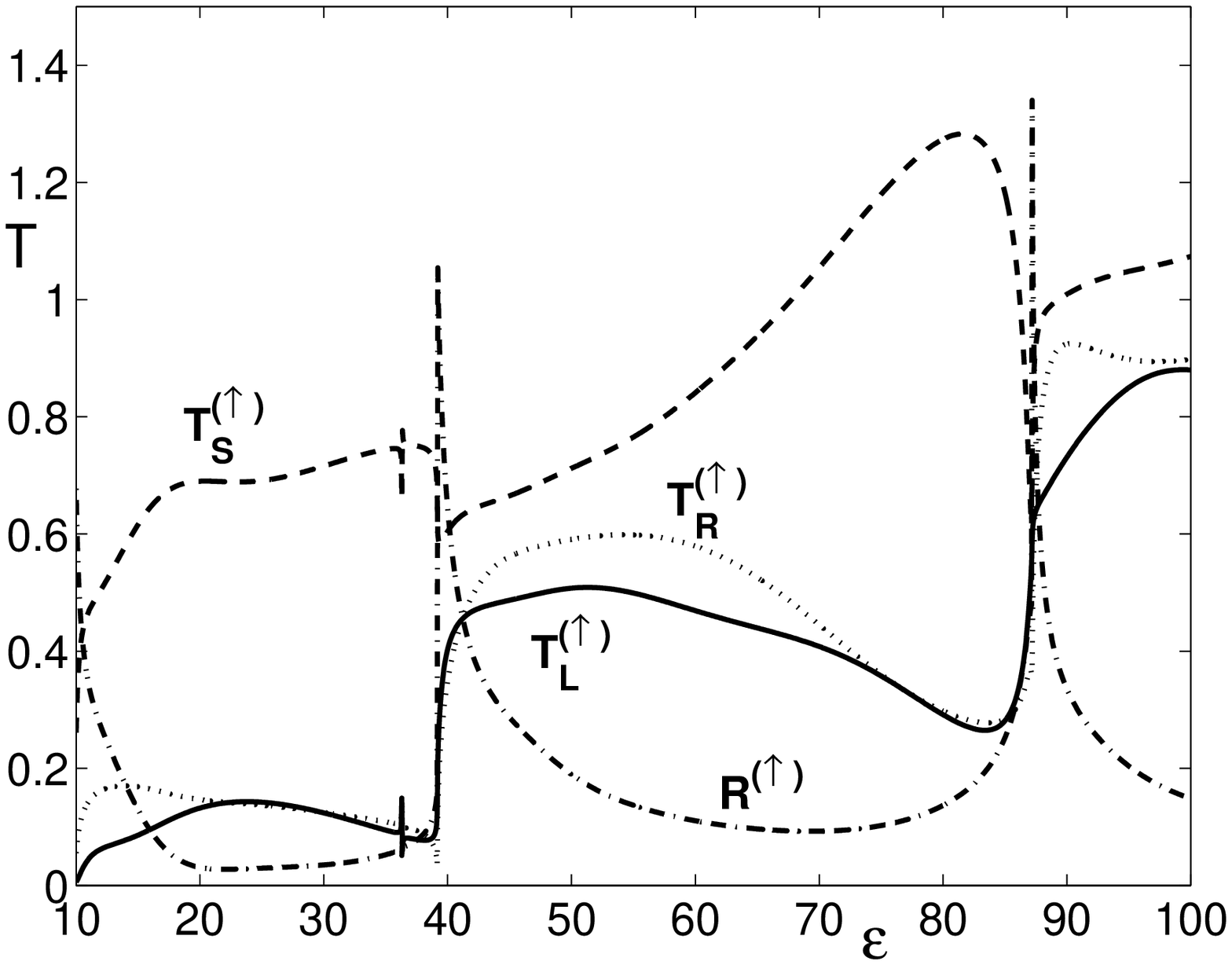}}
{3.1in}
{\footnotesize \noindent
FIG.3. Scattering coefficients $T^{(\uparrow)}_{L}$,
$T^{(\uparrow)}_{S}$, $T^{(\uparrow)}_{R}$ and $R^{(\uparrow)}$
for spin-up polarized electron wave injected  from the source
(lead 1) as function of the energy $\varepsilon$.}

\emptyplace{3.3in
\includegraphics{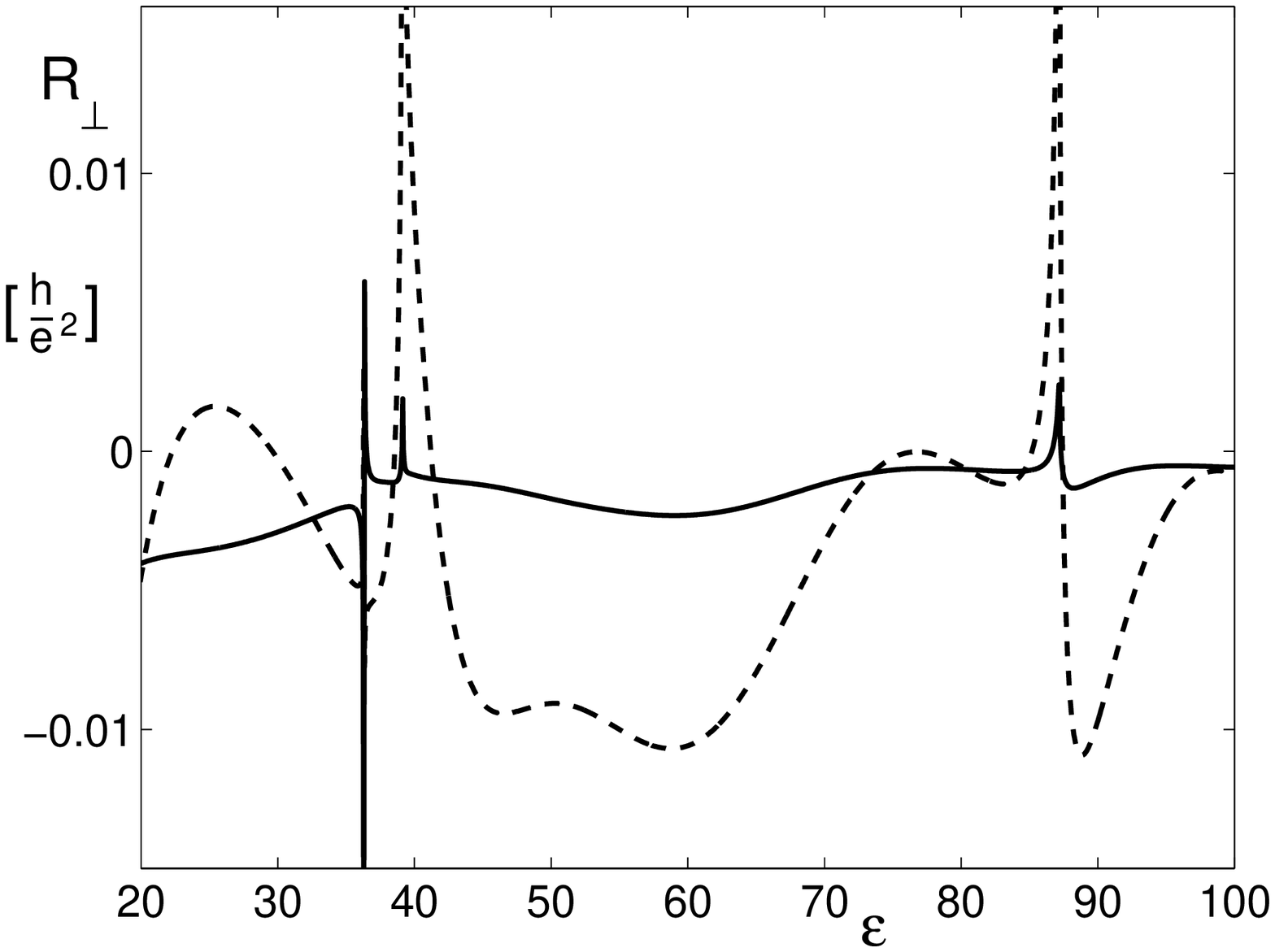}}
{3.1in}
{\footnotesize \noindent
FIG.4. Hall-like resistance induced by polarization filter
transparent for electron waves polarized along $\hat{z}$
direction (full line) and along $\hat{x}$ direction (dotted
line) as function of the energy.}

\title{Hall-like effect induced by spin-orbit interaction}
\author{E. N. Bulgakov$^{1}$,
K. N.Pichugin$^{1,2}$,
A. F. Sadreev$^{1,2}$}
\address{$1) \:$Kirensky Institute of Physics, 660036, Krasnoyarsk,
Russia \\
$2) \:$\AA bo Akademi, Institutionen f\"{o}r Fysik,
Department of Physics, SF-20500, \AA bo, Finland}

\author{P. St\v{r}eda and P. \v{S}eba}
\address{Institute of Physics, Academy of Sciences of the
Czech Republic, Cukrovarnick\'{a}
10, 162 53 Praha}

\maketitle

\date{today}

\begin{abstract}
The effect of spin-orbit interaction on electron transport
properties of a cross-junction structure is studied. It is shown that
it results in spin polarization of left and right outgoing
electron waves. Consequently, incoming electron wave of a proper
polarization induces voltage drop perpendicularly to the direct
current flow between source and drain of the considered
four-terminal cross-structure. The resulting Hall-like resistance
is estimated to be of the order of 10$^{-3}$-10$^{-2}$ $h/e^2$ for
technologically available structures. The effect becomes more
pronounced in the vicinity of resonances where Hall-like
resistance changes its sign as function of the Fermi energy.
\end{abstract}

\narrowtext

The spin-orbit interaction has polarization effect on particle
scattering processes \cite{davidov}. It is well known that
unpolarized beam of nucleons scattered by a zero-spin nuclei
becomes polarized. On the other side polarized incident beam results
in azimuthal asymmetry of the scattering process. These effects
have been observed long time ago in Stern-Gerlach experiments.
Similar effects might be expected for electron scattering
processes in microstructures which can be viewed as electron
waveguides.

Influence of the spin-orbit interaction on the electron
transport properties of mesoscopic systems attracts attention
of physicists since early 80's. At that time it has
been found that it is responsible for so called antilocalization
effect \cite{hikami}. Later,
spin-orbit interaction in devices of the Aharonov-Bohm geometry
has been systematically studied. In one-dimensional rings
it affects the sign of the persistent currents \cite{meir,entin}
and leads to a topological spin phase \cite{aronov}. These effects
originate in spin-orbit coupling term which is linear in momentum
$\vec{p}$
\begin{equation}
\label{(1)}
\frac{1}{2m^2} \sigma \times \vec{\nabla}V(\vec{r}) \cdot \vec{p}
\, \approx \,
\frac {\hbar}{2} \sum_{\mu,\nu} \sigma_{\mu} \beta_{\mu,\nu}p_{\nu},
\end{equation}
where $\sigma_{\mu}$ denotes Pauli matrices, $V(\vec{r})$ is a
background potential and $\beta_{\mu,\nu}$
represents a coupling strength.  This term is responsible
for spin-orbit splitting of electron states at $p\neq 0$.
Just recently splitting of Aharonov-Bohm oscillations caused by a
strong spin-orbit coupling has been reported \cite{morpurgo}.

In semiconductor-based devices there are two main contributions
to the spin-orbit coupling \cite{rashba1,rashba2}. One of them
arises due to absence of an inversion centre in the bulk
A$_{\rm III}$B$_{\rm V}$ material, from which devices are usually
fabricated, resulting in $k$-odd terms in the Hamiltonian of
3D-electrons. The second contribution originates in a low spatial
symmetry of the confining potential caused by asymmetry of the
space charge distribution. Lifting of the spin degeneracy in zero
magnetic field has already been
experimentally verified for two-dimensional electron systems in
different semiconductor structures \cite{das,doroz,nitta}.
In all cases the found spin-orbit coupling constant $\hbar^2 \beta$
has been of the order of few mV$\cdot$nm.

Anomalous resistance due to asymmetry of elastic scattering
processes induced by spin-orbit interaction might be expected
for the cross-junction device sketched in Fig.~1. Transmission
probabilities between perpendicular arms of the device should differ
for spin-up and spin-down states of incident electrons. Consequently
polarized incident electron beam may lead to Hall-like effect
in the absence of an external magnetic field.

\inspicture

We will assume
the cross-junction device fabricated from a semiconductor
heterostructure with a two-dimensional electron gas.
The model Hamiltonian of such systems is usually assumed to be
of the following form \cite{aronov,lyanda}
\begin{equation}
\label{soiham}
H \, = \, \frac{p_{x}^{2} + p_{y}^{2}}{2m^{*}} \, + \,
\hbar \beta \left [ \sigma_{x} p_{y} - \sigma_{y} p_{x} \right ]
\, + \, V(x,y) \;.
\end{equation}
where potential $V(x,y)$ represents hard-wall conditions at the
device boundary, i.e. it is zero inside and  infinite outside
of the cross-junction area. The coupling strength $\beta$
represents an effective electric field along $\hat{z}$ direction
given by the form of the confining potential and absence of an
inversion centre.

The Hamiltonian $H$ is invariant under time reversal
represented by the operator $\hat{T}= i \sigma_y K$
with  $K$ being  the operator of complex conjugation. The spin
matrix $i \sigma_y$ acting upon the wavefunction of a state
with well-defined value of the $z$-component of the spin, $s_z$,
changes the value of the $z$-component of the spin to its
opposite, $-s_z$ \cite{davidov}.

For the symmetrical cross-structure described by the Hamiltonian
$H$ there is
additional inversion symmetry related to transformation
$x \rightarrow -x$ and $y \rightarrow -y$ represented by
the operator $\hat{P}_x$ and $\hat{P}_y$, respectively.
The Hamiltonian $H$, Eq.(\ref{soiham}), commutes with operators
$\sigma_x \hat{P}_x$ and $\sigma_y \hat{P}_y$ and transformed
eigenfunctions
\begin{eqnarray}
\label{inv}
\psi'(x,y) \, & = & \, \sigma_y P_y\psi(x,y) \; ,
\nonumber \\
\psi'(x,y) \, & = & \, \sigma_x P_x\psi(x,y) \;
\end{eqnarray}
are thus eigenfunctions of the same Hamiltonian as well.

Current and voltage contacts are modelled by huge electron
reservoirs with negligible spin-orbit interaction.
To simplify scattering boundary condition we have placed ideal
leads with vanishing spin-orbit coupling, $\beta \equiv 0$,
between electron reservoirs and studied cross-junction,
as sketched in Fig.~1. In these asymptotic regions electron
wavefunctions can be expressed as linear combination of
eigenfunctions of the straight infinite lead at given energy
$E$. For each subband $n$ they have form of a plane wave, e.g.
for leads connecting reservoirs 1 and  3 we have
\begin{eqnarray}
\label{leadpsi}
\psi(x,y) \, & = & \, \sqrt{\frac{1}{\pi w}}
e^{\pm ik_n x} \sin \frac{\pi n y}{w} \chi(s_z)
\nonumber \\ \\
\chi(s_z) \, & = & \,
\left(\begin{array}{c} 1\\0 \end{array}\right)
{\rm or}
\left(\begin{array}{c} 0 \\ 1 \end{array}\right)
\nonumber
\end{eqnarray}
where $w$ denotes the lead width and
$k_n^2 = 2m^{\ast}E/\hbar^2 - \pi^2 n^2/w^2$.  Each spin state
$s_z=\pm${\scriptsize$\frac{1}{2}$},
within given subband $n$ is forming its own quantum channel.

Electron transport properties of a quantum device allowing elastic
scattering only are fully determined by transition probabilities
$t_{i,j}(n,s_z \rightarrow m,s'_z)$ representing electron
transition of the wave $k_n$ with spin $s_z$ approaching crossing
via $i$-th lead into an outgoing channel-state ($k_m,s'_z$)
within the lead $j$. The symmetry properties of the Hamiltonian
$H$ discussed above imply following useful identities
\begin{eqnarray}
\label{tsym}
t_{1,2}(n,s_z \rightarrow m, s'_z) \, & = & \,
t_{1,4}(n,- s_z \rightarrow m,- s'_z) \; ,
\nonumber \\
t_{1,2}(n,s_z \rightarrow m,- s_z) \, & = & \,
t_{1,2}(n,- s_z \rightarrow m, s_z) \; ,
\\
t_{1,3}(n,s_z \rightarrow m,s'_z) \, & =  & \,
t_{1,3}(n,- s_z \rightarrow m,-s'_z)   \; .
\nonumber
\end{eqnarray}
that remain valid for
cyclic interchange of the lead numbering.

To obtain transition probabilities the following coupled
equations for electron eigenfunctions have to be solved
\begin{eqnarray}
\label{equ}
\frac{{\partial}^2 u_1}{\partial{x^2}} \, + \,
\frac{{\partial}^2 u_1}{\partial{y^2}} \, + \, \varepsilon \,  u_1
\, + \, i \alpha \frac{{\partial}{u_2}}{\partial{y}}
\, - \, \alpha \frac{\partial u_2}{\partial{x}} \, = \, 0 \; ,
\nonumber \\  \\
\frac{{\partial  u_2}^2}{\partial{x^2}} \, + \,
\frac{{\partial}^2  u_2}{\partial{y^2}}  \, + \, \varepsilon \, u_2
\, + \, i\alpha \frac{{\partial}{u_1}}{\partial{y}} \, + \,
\alpha \frac{\partial u_1}{\partial{x}} \, = \, 0  \; ,
\nonumber
\end{eqnarray}
together with scattering boundary conditions discussed above.
Here we have introduced the following notations
\begin{displaymath}
\psi \, \equiv \, \left (
\begin{array}{l}
u_1 \\ u_2
\end{array}
\right ) \quad , \quad
\varepsilon = \frac{2m^{*}w^2 E}{\hbar^2} \quad , \quad
\alpha = 2m^{*} \beta w \; . \nonumber
\end{displaymath}
The value $\alpha=1$  has been chosen for numerical
calculation. It represents an InAs structure
($m^{*}=0.023 m_0$) of the lead width $w \approx$ 0.2 $\mu$m
and with the spin-orbit coupling constant
$\hbar^2 \beta \sim 6\times 10^{-3} {\rm eV} \cdot {\rm nm}$
\cite{lyanda}. Numerical results have been obtained by using
similar procedure as that already described by Ando \cite{ando}.

To describe scattering asymmetry for more general case of several
subbands it is useful to introduce partial transmission
coefficients
$T_{i \uparrow ,j \uparrow}$, $T_{i \uparrow ,j \downarrow}$,
$T_{i \downarrow ,j \uparrow}$ and $T_{i \downarrow ,j \downarrow}$
representing scattering of the fully polarized wave along $\hat{z}$
direction ($s_z = ${\scriptsize $\frac{1}{2}$} or
-{\scriptsize $\frac{1}{2}$}) into outgoing channels
of one particular spin orientation.
They are given as the sum of transition probabilities, Eqs.(\ref{tsym}),
over relevant channels, well defined in asymptotic lead-regions.
Obtained  spin-depend  coefficients representing electron transitions
from the lead 1 into the left arm of the cross-junction are shown
in Fig.~2. Partial transmission coefficients representing right
turn have the same energy dependence and are related to those
describing left turn as follows
\begin{eqnarray}
\label{parTsym}
T_{1 \uparrow ,2 \uparrow} \; = \; T_{1 \downarrow ,4 \downarrow}
\; \; & , & \; \;
T_{1 \downarrow ,2 \downarrow} \; = \; T_{1 \uparrow ,4 \uparrow}
\; \; ,
\nonumber \\
T_{1 \uparrow ,2 \downarrow} \; = \; T_{1 \downarrow ,4 \uparrow}
\; \; & = & \; \;
T_{1 \downarrow ,2 \uparrow} \; = \; T_{1 \uparrow ,4 \downarrow}
\; \; .
\end{eqnarray}
These identities are a direct consequence of the symmetry of
transition probabilities, Eg.(\ref{tsym}).

\inspicture

It is natural to suppose that reservoirs act as black bodies and
that they are emitting and absorbing electrons independently on
their spin orientation. It implies that incoming wave should be
considered as an unpolarized wave. Nevertheless, even in this case
left and right outgoing waves can be partially polarized since
the transmission into spin-up channels
($T_{1 \uparrow ,2 \uparrow} + T_{1 \downarrow ,2 \uparrow}$)
differs from that into spin-down channels
($T_{1 \uparrow ,2 \downarrow} + T_{1 \downarrow ,2 \downarrow}$),
as seen in Fig.~2.

Asymmetry of the scattering process also leads to the
tendency of injected electrons to prefer left or right turn
at the crossing in the dependence on their spin orientation.
To study this effect we have evaluated scattering coefficients
$T^{(\uparrow)}$ and $T^{(\downarrow)}$ for the case of fully
polarized incoming-waves defined as follows
\begin{eqnarray}
\label{Tsym}
T^{(\uparrow)}_{L} & \equiv &
T_{1 \uparrow ,2 \uparrow} + T_{1 \uparrow ,2 \downarrow}
\, = \,
T_{1 \downarrow ,4 \uparrow} + T_{1 \downarrow ,4 \downarrow}
\, = \, T^{(\downarrow)}_{R} \; , \; \nonumber \\
T^{(\uparrow)}_{S} & \equiv &
T_{1 \uparrow ,3 \uparrow} + T_{1 \uparrow ,3 \downarrow}
\, = \,
T_{1 \downarrow ,3 \uparrow}  + T_{1 \downarrow ,3 \downarrow}
\, = \, T^{(\downarrow)}_{S} \; , \; \\
T^{(\uparrow)}_{R} & \equiv &
T_{1 \uparrow ,4 \uparrow} + T_{1 \uparrow ,4 \downarrow}
\, = \,
T_{1 \downarrow ,2 \uparrow} + T_{1 \downarrow ,2 \downarrow}
\, = \, T^{(\downarrow)}_{L} \; , \; \nonumber \\
R^{(\uparrow)} & \equiv &
R_{i \uparrow ,i \uparrow} + R_{i \uparrow ,i \downarrow}
\, = \,
R_{i \downarrow ,i \uparrow} + R_{i \downarrow ,i \downarrow}
\, = \, R^{(\downarrow)} \; , \; \nonumber
\end{eqnarray}
Their energy dependence is shown in Fig.~3.
Other set of identities can be obtained by cyclic
interchange of the lead numbering.

\inspicture

Expected tendency of electrons with one particular spin
orientation to prefer left or right turn is evident.
Exceptions have been found in the
vicinity of subband edges at energies  $\varepsilon_n = \pi^2 n^2$.
Sharp peak in the transmission probabilities also appear at the energy
of the second bound state ($\varepsilon_b \approx 36.72$) formed in
cross structures \cite{schult}. It originates in a mixing of
bound and transport states caused by spin-orbit interaction. It
is similar effect as that induced by radiation field \cite{bulgakov}.

Under particular conditions the discussed spin dependent
scattering could lead to a Hall-like effect.
Current flow J applied along $\hat{x}$ direction, i.e. from a
source $1$ to a drain $3$, could not only induce a voltage
drop between source and drain, $U_{\|} \equiv U_1 - U_3$, but
there might also appear a voltage drop in perpendicular direction,
between voltage contacts $2$ and $4$, $U_{\perp} \equiv U_2-U_4$.
Their relation can be expressed with the help of reflection
coefficients ${\cal R}_{ii}$ and transmission coefficients
${\cal T}_{ij}$ representing electron transition from the
contact $i$ to the contact $j$\cite{buttiker}. For the considered
four terminal device we get
\begin{equation}
\label{hall}
U_{\perp} \; = \;
\frac{{\cal T}_{12}{\cal T}_{34} \, - \,
{\cal T}_{14}{\cal T}_{32}}
{(2N - {\cal R}_{22})(2N - {\cal R}_{44}) \, - \,
{\cal T}_{24}{\cal T}_{42}} \, U_{\|} \; \; ,
\end{equation}
where $N$ denotes number of the subbands at given energy.
To get non-zero value of $U_{\perp}$ it is necessary to
inject a polarized electron wave into the cross-junction
device. To ensure it let us place a filter into the lead $1$
which is assumed to be fully transparent for spin-up electrons,
$s_z=${\scriptsize$\frac{1}{2}$}. Spin-down electrons are
supposed to be reflected by the filter.
Only injected spin-up electrons are thus allowed to
reach the region between filter and
crossing denoted in Fig.1 as $1'$. Those spin-up electrons
that are reflected back by crossing into a spin-down channel are
not allowed to leave region $1'$ immediately. They are reflected
by filter and can try to escape from region $1'$ again. The
followed multiple-reflection process is controlled by reflection
coefficient $R_{1\downarrow,1\downarrow}$.
For transmission coefficients entering numerator of the right
hand side of Eq.(\ref{hall}) we get:
\begin{eqnarray}
{\cal T}_{12} & = &
T^{(\uparrow)}_{L} + T^{(\downarrow)}_{L}
- \gamma_1 T^{(\downarrow)}_{L}  \; ,
\nonumber \\
{\cal T}_{14} &  = &
T^{(\uparrow)}_{R} + T^{(\downarrow)}_{R}
- \gamma_1 T^{(\downarrow)}_{R}  \;,
\\
{\cal T}_{32} & = &
T^{(\uparrow)}_{R} + T^{(\downarrow)}_{R}
+ \gamma_3 T^{(\downarrow)}_{L}  \; ,
\nonumber \\
{\cal T}_{34} & = &
T^{(\uparrow)}_{L} + T^{(\downarrow)}_{L}
+ \gamma_3 T^{(\downarrow)}_{R}  \; .
\nonumber
\end{eqnarray}
Coefficients $\gamma_i$ represent effect of the filter
and they have the following form
\begin{eqnarray}
\label{gamma}
\gamma_1  =  \frac{T^{(\downarrow)}_{R} +
T^{(\downarrow)}_{S} + T^{(\downarrow)}_{L}}
{N-R_{1\downarrow,1\downarrow}}
\; ; \;
\gamma_{3}  =
\frac{T_{3\uparrow,1 \downarrow} + T_{3 \downarrow ,1 \downarrow}}
{N - R_{1 \downarrow, 1 \downarrow}}
\; .
\end{eqnarray}
For simplicity we have assumed that there is no spin-flip process
associated with reflections at the filter boundary. We have also
neglected any interference effects due to multiple scattering
processes in the region 1' between filter and crossing assuming
an inelastic equilibration processes in the filter vicinity leading
to equal occupation of spin-down channels.

Inserting expressions for scattering coefficients ${\cal R}_{ii}$
and ${\cal T}_{ij}$
into Eq.(\ref{hall}) and making use of the symmetry relations,
Eq.(\ref{parTsym}) and Eq.(\ref{Tsym}), we get
\begin{equation}
U_{\perp} =
\frac{\frac{1}{4}
\left ( T^{(\uparrow)}_{L} - T^{(\uparrow)}_{R} \right )}
{N - R_{1 \downarrow, 1 \downarrow} - \frac{1}{2}
\frac{T^{(\uparrow)}_{S}(T^{(\uparrow)}_{R}+T^{(\uparrow)}_{L})
+2T^{(\uparrow)}_{R}T^{(\uparrow)}_{L}}
{T^{(\uparrow)}_{R}+2T^{(\uparrow)}_{S}+T^{(\uparrow)}_{L}}}
\, U_{\|} .
\end{equation}
The voltage $U_{\perp}$ is proportional to the difference
$T^{(\uparrow)}_{L}-T^{(\uparrow)}_{R} \equiv
T^{(\downarrow)}_{R}-T^{(\downarrow)}_{L}$ similarly as in the case
of the standard Hall resistance \cite{buttiker}.  Because of the
spin-orbit coupling this difference might be non-zero as can be
seen in Fig.~3. Without presence of the polarization filter no
perpendicular voltage drop arises as can be easily shown by
assigning zero values to the coefficients $\gamma_i$.
To obtain Hall-like resistance $R_{\perp} = U_{\perp} / J$ all
other scattering coefficients have to be evaluated. The numerical
results for the parameters already used to evaluate partial
transmission coefficients are shown in Fig.~4.

\inspicture

Two other types of polarization filters have also be considered.
We have found that no Hall-like resistance appears if
polarization filter is transparent for waves polarized along
$\hat{y}$ direction. On the other side the filter which is
transparent for electron waves polarized along the current
direction, $\hat{x}$ direction, Hall-like resistance becomes
even larger as shown in Fig.~4.

In the vicinity of bound states and subband edges the Hall-like
effect become stronger and changes its sign. It indicates
that there appear circulating currents in the crossing region
changing their orientation with energy. Their origin in the vicinity
of the second bound state with energy $\varepsilon_b \approx 36.72$
is understandable. Due to spin-orbit coupling this state is splitted
into two states with opposite orbital momentum similarly like in
devices of the Aharonov-Bohm geometry \cite{meir,entin}. Splitting of
subband edges has similar effect. However, for most energies
perpendicular voltage appears due to deformed current lines
within cross junction only.
Results of the model calculation slightly depend on
the position of the boundaries between regions with turn on and
turn off spin-orbit coupling. However, no qualitative changes
have been observed.

The obtained values of the Hall-like resistance are measurable.
However, available real cross-junctions are of larger dimensions
than that used in our calculation. For this reason we have
calculated $R_{\perp}$ till energies $\varepsilon$ four times
larger than that
presented in Fig.~4. As expected with increasing number of
channels the effect decreases but $R_{\perp}$ is still of the
order 10$^{-3} h/ e^2$. Polarization filters might be realized by
a locally applied magnetic field across the cross-junction arm,
e.g. making use of a ferromagnetic top layer. Also a ferromagnetic
injector \cite{datta} can be use to induce Hall-like voltage.
Polarization effects in real systems  will be
hardly so effective as supposed in our model calculation.
Especially possible spin-flip processes in the vicinity of spin
reflecting boundaries or due to imperfections within
device leads would  partially suppress the described Hall-like
effect. Nevertheless, the asymmetry induced by spin-orbit
coupling has been long time ago verified in particle scattering
experiments and we believe that the
discussed effects will be ones observed in nanostructures as well.

\acknowledgements
This work was supported by the Grant Agency of the Czech Republic
under Grant No. 202/98/0085 and by the Academy of Sciences of the
Czech Republic under Grant No. A1048804.
It has been also partially supported by the INTAS-RFBR Grant 95-IN-RU-657,
RFFI Grant 97-02-16305, Krasnoyarsk Science Foundation Grant No.
7F0130 and
by the "Foundation for Theoretical Physics" in Slemeno, Czech Republic.
Authors acknowledge simulating discussions with Pavel Exner.

\end{document}